\documentclass[dvips]{article}
\usepackage{graphicx}
\def\<{\langle}
\def\>{\rangle}

\def\bv{{\bf b}}

\def\Xv{{\bf X}}
\def\ev{{\bf e}}

\def\Av{{\bf A}}
\def\Bv{{\bf B}}
\def\Ev{{\bf E}}
\def\Vv{{\bf V}}

\def\xu{\hat{\bf x}}
\def\yu{\hat{\bf y}}
\def\zu{\hat{\bf z}}

\def\be{\begin{equation}}
\def\ee{\end{equation}}

\begin{document}
%
\Large
\centerline{PHOTON PRODUCTION BY CHARGED}
\centerline{PARTICLES IN NARROW OPTICAL FIBERS\footnote{
Presented at {\it International Conference on Charged and Neutral Particles Channeling Phenomena}, Frascati, Italy, July 3-7, 2006. 
}}
\normalsize
\vskip 0.6 true cm

\centerline{Xavier ARTRU%
\footnote{e-mail: x.artru@ipnl.in2p3.fr}
C\'edric Ray%
\footnote{e-mail: c.ray@ipnl.in2p3.fr}
}
\centerline{Institut de Physique Nucl\'eaire de Lyon,} 
\centerline{Universit\'e Claude-Bernard \& IN2P3-CNRS,} 
\centerline{69622 Villeurbanne, France}

\vskip 0.8 true cm
\centerline{\bf ABSTRACT}

\medskip\noindent
A charged particle passing through or by an optical fiber induces emission of light guided by the fiber. The formula giving the spontaneous emission amplitude are given in the general case when the particle trajectory is not parallel to the fiber axis. At small angle, the photon yield grows like the inverse power of the angle and in the parallel limiting case the fiber Cherenkov effect studied by Bogdankevich and Bolotovskii is recovered. Possible application to beam diagnostics are discussed, as well as resonance effects when the particle trajectory or the fiber is bent periodically. 

\medskip\noindent
{\bf keywords}
optical fibers, monomode fibers, cherenkov radiation, transition radiation, beam diagnostics

\section{Introduction}

A charged particle passing through an optical fiber of large enough ($a\gg\lambda$) radius produces Cherenkov radiation (CR) inside the fiber if its velocity is above the Cherenkov threshold $1/\sqrt{\varepsilon}$. Photons which are emitted at sufficiently small angle ($\cos\theta\ge1/\sqrt{\varepsilon}$) with the fiber axis undergo total reflection and can be guided by the fiber toward a photon detector, making a position sensitive particle detector \cite{DIRC}. 

Transition Radiation (TR), emitted when the particle crosses the fiber surface, is another mechanism producing guided light in the fiber. Such a mechanism has been suggested to produce X-rays in capillaries (``channeled transition radiation'' \cite{Zhevago}). It has no threshold in velocity. Above the Cherenkov threshold, adding naively  Cherenkov and Transition Radiations makes double counting. Indeed, Transition Radiation from the entrance and exit faces of a dielectric slab, added coherently, contains Cherenkov radiation as a singular part. Only for broad enough fibers can one apply the Cherenkov formula, neglecting the edge effects. 

The Transition Radiation formula, derived for the ideal case of an infinite flat boundary, can be applied to the case of a lamellar optical guide. However it is inadequate for narrow ($a\sim\lambda$) cylindrical fibers which have only one or a few number of modes, because of the strong curvature of the surface.
It is precisely this case ($a\sim\lambda$) that we will study here. 

\begin{figure}[htbp] 
	\centering
\includegraphics*[angle=-90, scale=.4,clip,bb=30 120 560 720]{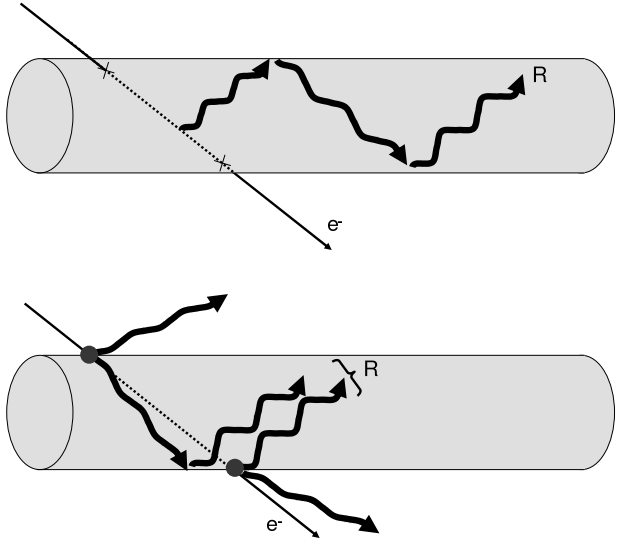}		
	\label{fig:FIG1}
\caption{Standard mechanisms of light emission. Top: Cherenkov radiation. Bottom: Transition radiation}
\end{figure}

Photon emission inside the fiber is also like the {\it cavity radiation} considered by Nitta and Miyazaki \cite{NittaMiya}. They consider rectangular cavities of finite size in all dimensions, where the frequency spectrum is discrete. In our case we have one quasi-infinite dimension and continuous spectrum. 

If the particle passes close to the fiber without touching it, its Coulomb field can be diffracted, giving rise to Diffraction Radiation. This radiation is mainly outside, i.e. in the ``free'' modes of the fiber, but nothing forbids some part to be kept ``inside''. This phenomenon will also be treated on the same footing as the case where the particle crosses the fiber.

\section{Basic radiation formula}

We take the fiber axis as $\zu$ axis. 
The quantized electromagnetic field $\Av^{op}$ in absence of the particle but in presence of the fiber can be expanded in proper modes of the longitudinal momentum $k_z$ (continuous), total  (orbital + spin) angular momentum $M\equiv L_z+S_z$ (discrete) about the $z$-axis and radial quantum number $\nu$ (continuous or discrete):   
$$
\Av^{op}(x,y,z,t)=\int^{\infty}_{-\infty} \frac{dk_z}{2\pi}
\sum^{\infty}_{M=-\infty} \sum^{\infty}_{\nu=0} 
a_{M,k_z,\nu} \ e^{izk_z-i\omega t}  \  \Av^{(M,k_z,\nu)} (x,y). 
$$
The radiation gauge $A^0=0$ has been chosen. We have quantized the field in an infinitely long cylindrical box, so that $k_z$ is continuous and $\nu$ is discrete not only for guided modes, but also for free ones.
The photon energy $\omega$ depends on $k_z$, $M$ and $\nu$ through a dispersion formula
\begin{equation}
\omega=\omega(M,k_z,\nu).
\label{disp}
\end{equation} 
Fig.2 shows the phase velocity $v_\phi = \omega/k_z$ of the lowest mode ($M=\pm1, \nu=1$) called $HE_{11}$ and the external fraction of the mode power, as a function of $\omega$. 
\begin{figure}[htbp] 
	\centering
\includegraphics*[scale=1.,clip,bb=10 10 330 230]{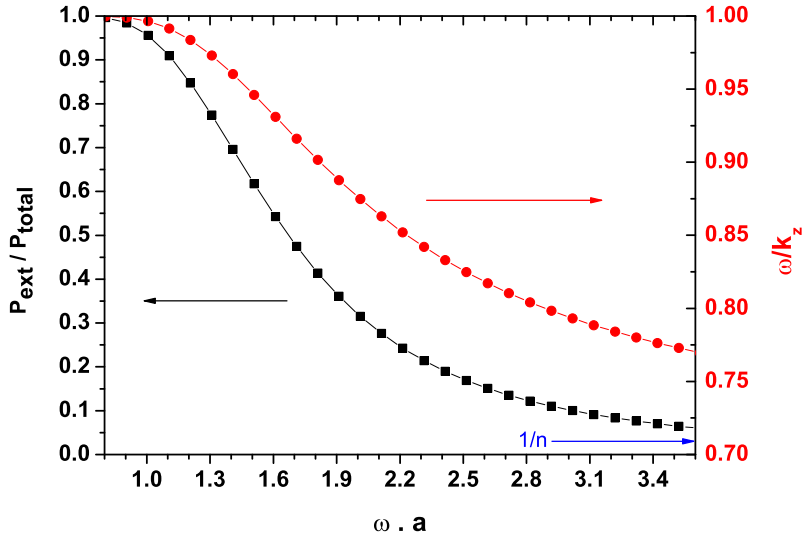}		
	\label{fig:FIG2}
\caption{phase velocity $v_\phi = \omega/k_z$ and external fraction of the power for the $HE_{11}$ mode.}
\end{figure}

The time-independent annihilation and creation operators obey the commutation rules
\begin{equation}
\left[a_{M,k_z,\nu} , a^\dagger_{M',k'_z,\nu'} \right] =
2\pi\delta(k_z-k'_z) \ \delta_{MM'} \ \delta_{\nu\nu'}.
\label{commut}
\end{equation}

A particle of charge $e$ is {\it crossing} or {\it passing by} the fiber along the (straight or curved) trajectoty $\Xv(t)$. The amplitude $R$ of spontaneous emission in the mode $k_z,M,\nu$ is given by

\begin{eqnarray}
R^*_{M,k_z,\nu} &=& e\int^{t=\infty}_{t=-\infty} d\Xv(t)\cdot \left[\Av_{M,k_z,\nu} (x,y) 
\ e^{izk_z-i\omega t}\right] 
\\
&=& {e\over i\omega} \int d\Xv\cdot \Ev_{M,k_z,\nu} (x,y) \ e^{izk_z-i\omega t} 
\label{ampli}
\end{eqnarray}
where $\Ev = i \omega \Av$ is the associated electric field, according to $\Ev = - \partial_t \Av - \nabla A_0$. The resulting photon distribution is 
\begin{equation}
dN_{M,k_z,\nu} = \frac{dk_z}{2\pi} \ \left| R_{M,k_z,\nu} \right|^2
\label{distrib}
\end{equation}

\section{\bf The field modes}

The fiber we consider consists in a dielectric rod of homogeous index surrounded by vacuum (no clading). A guided mode is such that
$$
\omega \le k_z \le \omega\sqrt{\varepsilon} \equiv \omega \, n.
$$
Outside the fiber, it has an evanescent wave $\sim e^{-\kappa r}$ corresponding to an imaginary transverse momentum $i\kappa$. Inside the fiber, the photon has real transverse momentum $q$. $q$ and $\kappa$ are given by
$$
q(M,k_z,\nu) = \left\{ \varepsilon\omega^2 - k_z^2 \right\}^{1/2}
\qquad
\kappa(M,k_z,\nu) = \left\{ k_z^2 - \omega^2 \right\}^{1/2}.
$$ 
The longitudinal components $E_z$ and $B_z$ of the electric and magnetic fields have longitudinal spin $S_z=0$ so that their orbital angular momentum $L_z=M$. Using cylindrical coordinates $(z,r,\phi)$ we can  therefore write
$$
E_z^{(M,k_z,\nu)} (x,y) = i\ e^{iM\phi} \ f_z (r),
\qquad
B_z^{(M,k_z,\nu)} (x,y) = e^{iM\phi} \ h_z (r),
$$
Inside each homegeneous region (i.e., core and vacuum) $f_z$ and $h_z$ obey the same differential equation
\begin{equation}
\left[ \partial_r^2 + {1\over r}\partial_r - {M^2\over r^2} + k_T^2(r) \right] \ f_z \ {\rm or} \ h_z = 0
\label{eq-rad}
\end{equation}
where
$$
k_T^2(r) \equiv \omega^2\epsilon(r) - k_z^2 \ = \quad
q^2 \ {\rm inside,} \quad -\kappa^2 \ {\rm outside,} 
$$
$$
\epsilon(r) = \quad \varepsilon \ {\rm inside,} \quad 1 \ {\rm outside.} 
$$
The solutions of (\ref{eq-rad}) are the Bessel functions $J_M$ and $K_M$. We have the limiting conditions:
\begin{itemize}
	\item $f_z$ and $h_z$ are regular at $r=0$. Therefore $f_z \propto h_z \propto J_M(qr)$ for $r\le a$,
	\item $f_z$ and $h_z$ are continuous at $r=a$,
	\item $f_z$ and $h_z$ vanish at $r\to\infty$. Therefore $f_z \propto h_z \propto K_M(\kappa r)$ for $r\le a$.
\end{itemize}
It follows that $f_z$ and $h_z$ are proportional to each other (it would not be the case if there were more than one boundary) and we can write:
\begin{equation}
f_z(r) = c_E \ \psi(r), \qquad  h_z(r) = c_B \ \psi(r), 
\label{fz-hz}
\end{equation}
$$
\psi(r) = J_M(qr) \ {\rm inside},  \qquad = c_K \ K_M(\kappa r) \ {\rm outside},
\qquad c_K = \frac{J_M(qa)}{K_M(\kappa a)} 
$$
$\psi$ is continuous at $r=a$, but not $\psi'$. So (\ref{eq-rad}) cannot be treated as an ordinary Schr\o dinger equation. 

The transverse components $\Ev_T$ and $\Bv_T$ can be expressed 
in terms of the radial and azimutal vectors, $\ev^{r} = \cos\phi \ \xu + \sin\phi \ \yu$ 
and $\ev^{\phi} = \cos\phi \ \yu - \sin\phi \ \xu$, times $L_z=M$ radial functions:
$$
\Ev_T = e^{iM\phi}  \ \left( f_r (r)  \ \ev^{r} + f_\phi (r)  \ \ev^{\phi} \right)
$$
$$
\Bv_T = e^{iM\phi} \ \left( h_r (r)  \ \ev^{r} + h_\phi (r)  \ \ev^{\phi} \right)
$$
They can also be expressed
in terms of the $S_z=\pm1$ eigenvectors $\ev^\pm = (\xu \pm i \yu)/2$, times $L_z=M\mp1$ radial functions:
\begin{eqnarray}
\Ev_T = e^{i(M-1)\phi} \ f_- (r)  \ \ev^+  +
e^{i(M+1)\phi} \ f_+ (r)  \ \ev^- 
\\
i \, \Bv_T = e^{i(M-1)\phi} \ h_- (r) \ \ev^+  + 
e^{i(M+1)\phi} \ h_+ (r) \ \ev^- 
\label{fh+-}
\end{eqnarray}
with
$$
f_\pm = f_r \pm i f_\phi \ ,
\qquad
-i \, h_\pm = h_r \pm i h_\phi
$$ 
The Maxwell equations lead to (see Appendix):
$$
f_\pm = \frac{-k_z \ c_E \pm \omega \ c_B}{k_T^2 (r)} \ \left( \psi' \mp M {\psi\over r} \right)
$$
$$
h_\pm = \frac{-k_z \ c_B \pm \omega\epsilon(r) \ c_E }{k_T^2 (r)} \ \left( \psi' \mp M {\psi\over r} \right)
$$
Owing to a property of the Bessel functions,
$$
J'_M(x) \pm {M\over x} J_M(x) = \pm J_{M\mp 1}(x) 
\qquad
K'_M(x) \pm {M\over x} K_M(x) = - K_{M\mp 1}(x), 
$$
one obtains the final expressions 
\begin{eqnarray}
f_\pm(r) = c^\pm_{fJ} \ J_{M\pm1}(qr) \ (r\le a), \qquad  c^\pm_{fK} \ K_{M\pm1}(\kappa r) \ (r > a),
\label{f+-=}
\\
h_\pm(r) = c^\pm_{hJ} \ J_{M\pm1}(qr) \ (r\le a), \qquad  c^\pm_{hK} \ K_{M\pm1}(\kappa r) \ (r > a),
\end{eqnarray}
with 
$$
 c^\pm_{fJ} = \mp \frac{-k_z \ c_E \pm \omega \ c_B}{q},
\qquad 
 c^\pm_{fK} = \frac{-k_z \ c_E \pm \omega \ c_B}{\kappa} \ c_K ,
$$
$$
c^\pm_{hJ} = \mp \frac{-k_z \ c_B \pm \omega\varepsilon \ c_E }{q},
\qquad 
 c^\pm_{hK} = \frac{-k_z \ c_B \pm \omega \ c_E }{\kappa} \ c_K .
$$

The continuity of $h_z$, $h_\phi$, $f_z$, $f_\phi$ and $\epsilon(r) f_r$ at the fiber surface leads to 
$$
\frac{c_B}{c_E} = - M Q \left[{J'_M(u) \over uJ_M(u)}+ {K'_M(w) \over wK_M(w) }\right]^{-1} 
= - {1\over MQ} \left[{\varepsilon J'_M(u) \over uJ_M(u)}+ {K'_M(w) \over wK_M(w) }\right] 
$$
where $u\equiv qa$, $w\equiv \kappa a$ and
$$
Q = \left[ {1 \over u^2}+ {1 \over w^2} \right] {k_z\over\omega}
= \left[ {\varepsilon \over u^2}+ {1 \over w^2} \right] {\omega\over k_z} 
$$
from the two expressions of $c_B/c_E$ we obtain \cite{Okoshi}
\begin{equation}
\left[ \frac{J'_M(u)}{u J_M(u)} + \frac{K'_M(w)}{w K_M(w)} \right]
\cdot
\left[ \frac{\varepsilon J'_M(u)}{u J_M(u)} + \frac{K'_M(w)}{w K_M(w)} \right]
=
M^2 \ \left( {1\over u^2} +  {1\over w^2} \right) 
\cdot
\left( {\varepsilon\over u^2} +  {1\over w^2} \right) 
\label{valprop}
\end{equation}
This equation, together with 
\begin{equation}
u^2 + w^2 = (\varepsilon -1) \ \omega^2 a^2 
\quad {\rm and} \quad
u^2 + \varepsilon w^2 = (\varepsilon -1) \ k_z^2 a^2 
\label{uw}
\end{equation}
determines the dispersion relation (\ref{disp}) between $\omega$ and $k_z$.
 
For optical fibers used in telecommunications, $\varepsilon = \varepsilon_1/\varepsilon_2$ is the relative permitivity between the core and the clad, and is close to unity. Then $E_x$ and $E_y$ can be treated like independent scalar fields and $\varepsilon$ is replaced by 1 in (\ref{valprop}). In our case, we have no clad and $\varepsilon$ is large, for instance $\varepsilon=(1.41)^2\simeq2$ for fused silica. So we cannot make such an approximation.

\section{\bf Normalization of the mode}
 
The commutation relations (\ref{commut}) imply that the mode $\Av$ is normalized to unit linear photon density, i.e., a linear energy density
$$
\frac{dW}{dz}\{\Av\} = \omega.
$$
This fixes the absolute value of the coefficients $c_E$ and $c_B$. However the electromagnetic energy density has not a simple expression if there the medium is dispersive. On the other hand the {\it power} is given by the simple expression
$$
P \equiv \frac{dW}{dt}\{\Av\} = \int dx \ dy \ \Re \left\{\Ev^* \times \Bv\right\}_z.
$$
where $\Re \left\{\Ev^* \times \Bv\right\} $ is the Poynting vector. In terms of the functions introduced in \ref{fh+-},
\begin{eqnarray}
P = {1\over2} \int 2\pi r \ dr \ \Re \left\{ f^*_- (r) \ h_- (r) 
- f^*_+ (r) h_+ (r) \right\}
\\
= \pi \int_0^a  r \ dr  \left\{ c^-_{fJ} \, c^-_{hJ} \, J_{M-1}^2(qr) 
-  c^+_{fJ} \, c^+_{hJ} \, J_{M+1}^2(qr) \right\}
\nonumber
\\
+ \pi \int_a^\infty  r \ dr  \left\{ c^-_{fK} \, c^-_{hK} \, K_{M-1}^2(\kappa r) 
- c^+_{fK} \, c^+_{hK} \, K_{M+1}^2(\kappa r) \right\}
\label{Poynt}
\end{eqnarray}
Using the two expressions of the group velocity
$$ 
v_g = \frac{P}{dW/dz} = \frac{d\omega}{dk_z}  
$$ 
we can transform (\ref{distrib}) as follows
\begin{equation}
dN_{M,k_z,\nu} = \frac{d\omega}{2\pi v_g} \ \left| R_{M,k_z,\nu} \right|^2
= \frac{\omega \ d\omega}{2\pi P} \ \left| R_{M,k_z,\nu} \right|^2.
\label{distrib'}
\end{equation}
This formula is invariant under an overall rescaling of the fields $c_E$ and $c_B$ since both $P$ and $|R|^2$ are quadratic in the fields. 

\section{\bf The mode emission amplitude}

We assume that the particle follows the straight trajectory 
$$
\Xv=\bv+t\Vv, \qquad \bv=(b,0,0), \qquad \Vv=(0,V_T,V_z)
$$
The spontaneous emission $\{$amplitude$\}^*$ (\ref{ampli}) becomes
$$
R_{M,k_z,\nu}^* = {e\over i\omega} \int^{\infty}_{-\infty} dy
\left[ E_y^{M,k_z,\nu} (x,y) + {V_z\over V_T} E_z^{M,k_z,\nu} (x,y) \right]
\ e^{iy {V_z k_z-\omega \over V_T}} 
$$
$$
= {e\over2\omega} \int^{\infty}_{-\infty} dy \left[ 
e^{-i\phi} \ f_-(r) - e^{i\phi} \ f_+(r) + 2 {V_z\over V_T} f_z(r) \right]
\ e^{iM\phi + isy}
$$
One arrives at the real expression
\begin{eqnarray}
R_{M,k_z,\nu} = {e\over\omega} \int^{\infty}_{0} dy \ \{ \ 
\cos[sy+(M-1)\phi] \ f_-(r) - \cos[sy+(M+1)\phi] \ f_+(r) 
\nonumber
\\
+ 2 (V_z/V_T) \ \cos(sy+M\phi) \ f_z(r) \ \}
\label{ampli=}	
\end{eqnarray}
where $r = \sqrt{b^2 + y^2}$, $\phi = \tan^{-1}(y/b)$, $s = (V_z k_z-\omega)/V_T$; 
$f_+$, $f_-$ and $f_z$ are given in (\ref{fz-hz},\ref{f+-=}). 

For small transverse velocity, the third term of the integrand becomes large, although the longitudinal component $f_z$ is generally smaller than the transverse ones, and the photon spectrum can reach high values. On the other hand the amplitude is expected to decrease very fast when the dimensionless parameter $sa$ gets much larger than unity, because the integrand oscillates many times in the important field region $|y|\le a$. From these considerations, at small $V_T$ one expects an almost monochromatic peak of height $\propto (1/V_T)^2$ centered at frequencies such that $s=0$, i.e., the electron and the wave travel at the same longitudinal velocity:
\begin{equation}
\omega/k_z \equiv v_\phi = V_z
\label{Cher}
\end{equation} 
(``fiber Cherenkov condition''). The width $\Delta\omega$ of this peak is given by the condition 
$a\,\Delta s\simeq1$, where
\begin{equation}
a \, \Delta s = |\Delta\omega - V_z \Delta k_z)| \ {a\over V_T} = |1-V_\phi/v_g| \ {a \ \Delta\omega \over V_T} 
=  \frac{dv_\phi/v_\phi}{ d\omega/\omega} \ {a \ \Delta\omega \over V_T} = 1
\end{equation}
Given $v_z$, the Cherenkov frequency $\omega_{Ch}$ can be obtained from the upper curve of Fig.2. Also the inverse of $dv_\phi/(v_\phi \ d\omega)$ is the interval between $\omega_{Ch}$ and the intersection of the tangent to this curve and the horizontal axis.    
$\Delta\omega$ being proportional to $V_T$, after integration over $\omega$ one obtains a net photon number $\propto 1/V_T$. 
At small $V_T$, one approaches the case when the electron runs parallel to the fiber over a long distance $L$ \cite {Bolot}, with $L \sim a/V_T$. The spectrum becomes discrete, given by the ``fiber Cherenkov condition'' (\ref{Cher}).

\section{\bf Results and discussion}

\begin{figure}[htbp] 
	\centering
\includegraphics*[scale=1.,clip,bb=10 10 340 250]{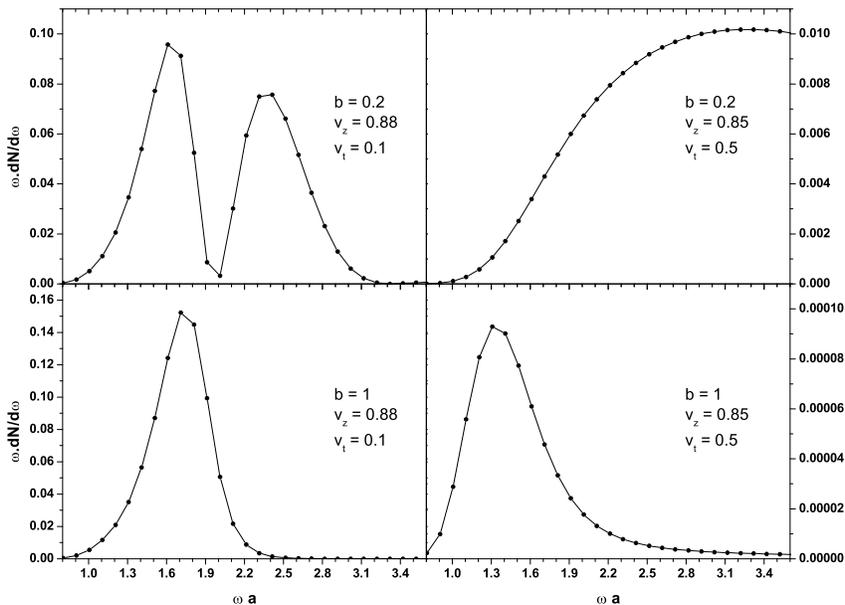}		
	\label{fig:FIG3}
\caption{Photon yields in the $HE_{11}$ mode for four types of the particle trajectory.}
\end{figure}

We have considered fibers of fused silica with n=1.41. The dimensionless photon yields $\omega dN / d\omega$ is plotted in Fig. 3 for two impact parameters, $b=0.2 \ a$ (penetrating trajectory) and $b=a$ (tangent trajectories), and two particle velocity vectors, $(v_z,v_T)=(0.88,0.1)$ (large angle) and $(v_z,v_T)=(0.85,0.5)$ (small angle). The angular momentum $M$ of the mode has been chosen to have the same sign as the particle's one. This is expected to be more favorable for radiation than the opposite sign case.

The spectrum from the penetrating trajectories are harder than from the tangent (or fully external) trajectories. This can be due to (i) the discontinuity of the fields at the fiber surface, (ii) the fact that the evanescent field is less important at high frequency. 

In the large angle - penetrating case, the yield is of the order of $\alpha = e^2/(4\pi) = 1/137$, as more or less expected. For the large angle - tangent case it is two orders of magnitude smaller. Note the peak at a relatively small frequency, where the wave travels mainly outside the fiber (see Fig.2). At still smaller frequency, the wave becomes too much diluted outside the fiber, which explains the vanishing yields in the four cases. 

Much bigger yields are obtained in the small angle cases, due to the increasing contribution of the longitudinal field discussed in the last section. Eq.(\ref{Cher}) should predict a peak at $\omega a \simeq2$. In the tangent case, this peak is shifted to the left, probably for the reasons (i) and (ii) said before.
In the penetrating case, instead of a peak, we have a dip at this very place. This is a peculiarity of the odd $M$ modes when $b$ is small. If $b=0$, then $\phi$ in (\ref{ampli=}) is either $-\pi/2$ or $+\pi/2$ and, at the Cherenkov point ($s=0$), $\cos(sy+M\phi)$ is zero in the whole integration range.

\section{Outlooks}

The production of coherent light by charged particle in a fiber may have applications in beam size measurement. Scintillating fibers yield more photons, but may be less hard to radiation than, e.g. fused silica. One can also compare with optical transition radiation and diffraction radiation devices. In our case, the synchrotron light from an upstream magnet or the diffraction radiation from a collimator does not make any background. It is just scattered, but not captured by the fiber, due to the orthogonality between the free and guided modes.

Unlike for optical transition radiation and diffraction radiation we do not expect a logarithmic increase at large Lorentz factor $\gamma$. Such an increase is usually due to the expansion of the photon cloud accompagnying the particle, but such photons are quasi-real, therefore almost orthogonal to the guided modes. 

Eq.(\ref{ampli}) can be applied to cases of bent particle trajectory or, using the adiabatic approximation for the modes, of bent fiber. Successive encounters will give rise to interference phenomena. 
A straight fiber inside an electron undulator will get filled by monochromatic radiation, $\omega$ and $k_z$ being related by the resonance condition
\begin{eqnarray}
	(k_z  - \omega / \left\langle V_z \right\rangle) \ \Lambda \equiv  
		k_z \Lambda - \omega L/V = 2m\pi 
\end{eqnarray}
where $\Lambda$ is the undulator period and $L = \Lambda V/\left\langle V_z \right\rangle$ the lenght of one period of trajectory. 
Similarly, a particle in straight motion close to a periodically bent fiber will induce monochromatic light in the fiber. This is a kind of {\it internal} Smith-Purcell radiation. 
If the fiber draws a {\it planar} periodic curve, the resonance condition is
\begin{eqnarray}
	k_L L - \omega \Lambda/V = 2m\pi 
\end{eqnarray}
where now $\Lambda$ is the spatial period of the fiber, $L$ its corresponding arc length and $k_L$
the wave number along the fiber. 
If the fiber draws an {\it helical} curve, the phase velocities of left- and right-handed polarisations in the fiber are split and the condition becomes
\begin{eqnarray}
	k_L L - \omega \Lambda/V = 2m\pi \pm \phi_B
\end{eqnarray}
where $\phi_B$, called {\it Berry phase}, is equal to the solid angle of the cone drawn by the tangent to the fiber \cite{TomitaChiao}.


\medskip
\begin{thebibliography}{99}
%
\bibitem{DIRC} P. Coyle et al, Nucl. Instr. Methods A 343 (1994) 292.
%
\bibitem{Zhevago} N.K. Zhevago, V.I. Glebov, Phys. Lett. A 309 (2003) 311; X. Artru, S.P. Fomin, N.F. Shul'ga, K.A. Ispirian, N.K. Zhevago, Phys. Reports 412 (2005) 89.  
%
\bibitem{NittaMiya} Phys. Rev. E 66 (2002) 035501.
%
\bibitem{Bolot} L.S. Bogdankevich, B.M. Bolotovskii, J. Exp. Theoret. Phys. 32, 1421 [Sov. Phys. JETP 5, 1157] (1957). See also N.K. Zhevago, V.I. Glebov, Nucl. Instr. Methods A 331 (1993) 592; Zh. Exp. Teor. Fiz. 111 (1997) 466. 
%
\bibitem{Okoshi} see, for instance, T. Okoshi {\it Optical Fibers}, Academic Press, 1982 
%
\bibitem{TomitaChiao} A. Tomita, R. Chiao, Phys. Rev. Lett. 57 (1986) 937.

\end{thebibliography}
\end{document}